\documentclass[twocolumn,twoside,slac_two]{revtex4}
\usepackage{graphicx}
\usepackage{fancyhdr}
\setcitestyle{square,numbers}
\def\epo{{\sc epos}}

\begin{document}

\title{New Developments of EPOS 2}

%
\author{T. Pierog}
\affiliation{KIT, Instit\"ut f\"ur Kernphysik, Karlsruhe, Germany}
\author{Iu. Karpenko}
\affiliation{Bogolyubov Institute for Theoretical Physics, Kiev, Ukraine}
\author{S. Porteboeuf}
\affiliation{University of Clermont-Ferrand, Clermont-Ferrand, France}
\author{K. Werner}
\affiliation{SUBATECH, University of Nantes -- IN2P3/CNRS-- EMN, Nantes,
France}

\begin{abstract}
Since 2006, \epo~hadronic interaction model is being used for very high energy 
cosmic ray analysis. Designed for minimum bias particle physics and used for having
a precise description of SPS and RHIC heavy ion collisions, \epo~brought 
more detailed description of hadronic interactions in air shower development. 
Thanks to this model it was possible to understand why there were less muons 
in air shower simulations than observed in real data. With the start of the 
LHC era, a better description of hard processes and collective effects is 
needed to deeply understand  the incoming data. We will describe the basic 
physics in \epo~and the new developments and constraints which are taken into 
account in \epo~2.

\end{abstract}

\maketitle

\thispagestyle{fancy}


\section{INTRODUCTION}

Air shower simulations are a very powerful tool to interpret ground
based cosmic ray experiments. However, most simulations are still
based on hadronic interaction models  more than 10 years old.
Much has been learned since, in particular due to new data available
from the SPS and RHIC accelerators. 

In this paper, we discuss the new development of the \epo~model, the
latter one being a hadronic interaction model, which does very well
compared to RHIC data~\cite{Bellwied}, and also other
particle physics experiments (especially SPS experiments at CERN). 
Used for air shower analysis since 2006, the last version 1.99 released 
in 2009 gives very interesting results in terms of mass 
composition~\cite{private}. Due to 
the constraints of particle physics, air shower simulations using \epo~present 
a larger number of muons at ground~\cite{Pierog:2006qv}. It allows for the first
time to reproduce both the muon number and the elongation rate using
a reasonable average mass for all energies between the knee and the 
Greisen-Zatsepin-Kuzmin (GZK) cut-off. 

On the other hand, the new measurements at LHC give us the opportunity to 
test the model more deeply. After a general introduction of the \epo~model,
we will explain in this paper the new developments done in \epo~2.

\section{EPOS MODEL}

One may consider the simple parton model to be the basis of high energy 
hadron-hadron interaction models, which can be seen as an exchange of a 
``parton ladder'' between the two hadrons.
\begin{figure}[th]
\begin{center}\includegraphics[width=0.24\textwidth]{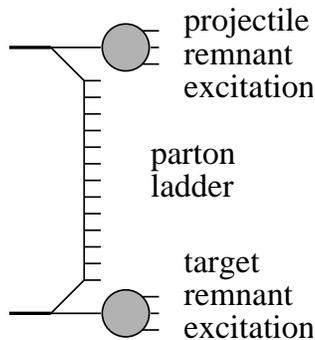}\end{center}
\caption{Elementary parton-parton scattering: the hard scattering in the
middle is preceded by parton emissions attached to remnants. The remnants 
are an
important source of particle production even at RHIC energies.\label{ladder} }
\end{figure}

In \epo, the term ``parton ladder'' is actually meant to contain two parts \cite{nexus}:
the hard one, as discussed above, and a soft one, which is a purely
phenomenological object, parameterized in Regge pole fashion.

In additions to the parton ladder, there is another source of particle production:
the two off-shell remnants, see fig. \ref{ladder}.
We showed in ref. \cite{nex-bar}
that this {}``three object picture'' can
solve the {}``multi-strange baryon problem'' of conventional
high energy models, see  ref. \cite{sbaryons}.

Hence \epo~is a consistent quantum mechanical multiple scattering approach
based on partons and strings~\cite{nexus}, where cross sections
and the particle production are calculated consistently, taking into
account energy conservation in both cases (unlike other models where
energy conservation is not considered for cross section calculations~\cite{hladik}).
Nuclear effects related to Cronin transverse
momentum broadening, parton saturation, and screening have been introduced
into \epo~\cite{splitting}. Furthermore, high density effects leading
to collective behavior in heavy ion collisions are also taken into
account~\cite{corona}.

\begin{figure}[htbp]
{\par \hfill
\begin{center}
\includegraphics[  scale=0.35]{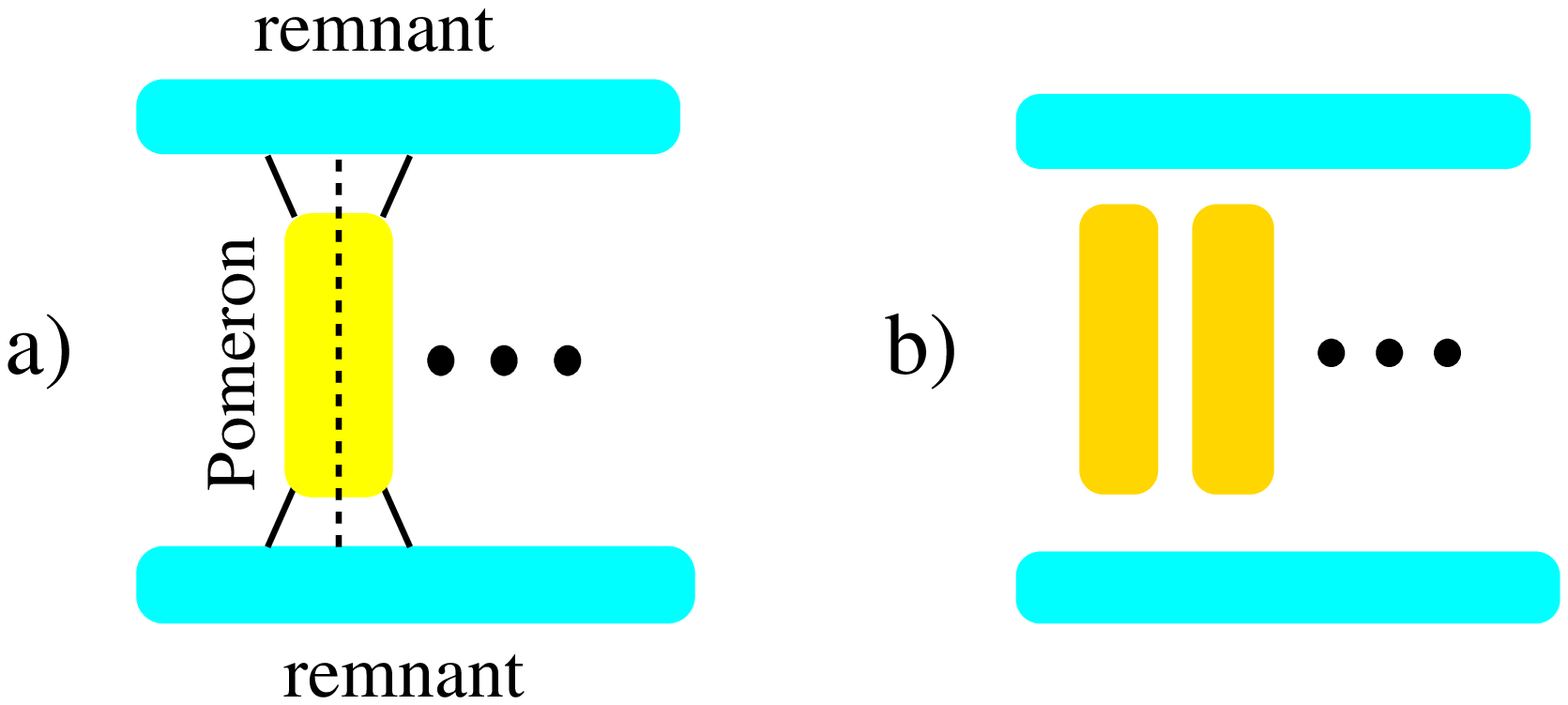}\hspace{1cm}
\includegraphics[  scale=0.45]{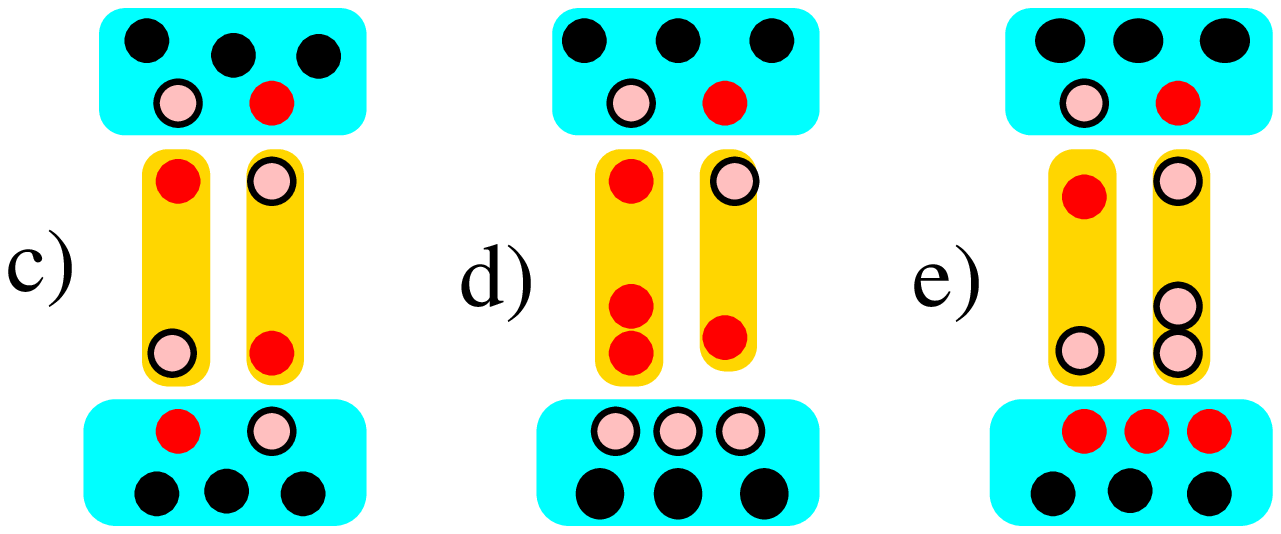}
\end{center}\hfill
\par}
\caption{ a) Each cut Pomeron is regarded as two strings b). 
c) The most simple and frequent collision
configuration has two remnants and only one cut Pomeron represented
by two $\mathrm{q}-\overline{\mathrm{q}}$ strings. d) One of the
$\overline{\mathrm{q}}$ string ends can be replaced by a $\mathrm{qq}$
string end. e) With the same probability, one of the $\mathrm{q}$
string ends can be replaced by a $\overline{\mathrm{q}}\overline{\mathrm{q}}$
string end. 
\label{nexus1}}
\end{figure}

Thanks to a Monte Carlo, first the collision configuration is determined:
i.e. the number of each type of Pomerons exchanged between the projectile
and target is fixed and the initial energy is shared between the Pomerons
and the two remnants. Then particle production is accounted from two
kinds of sources, remnant decay and cut Pomeron. A Pomeron may be
regarded as a two-layer (soft) parton ladder attached to projectile
and target remnants through its two legs. Each leg is a color singlet,
of type q$\overline{\mathrm{q}}$ , qqq or 
$\overline{\mathrm{q}}$$\overline{\mathrm{q}}$$\overline{\mathrm{q}}$ from the
sea, 
and then each cut Pomeron is regarded as two strings, Cf. fig.~\ref{nexus1}~a) 
and b). %

It is a natural idea to take quarks and anti-quarks from the sea as
string ends for soft Pomeron in \epo, because
an arbitrary number of Pomerons may be involved. 

\begin{figure}
\begin{center}\includegraphics[width=0.47\textwidth]{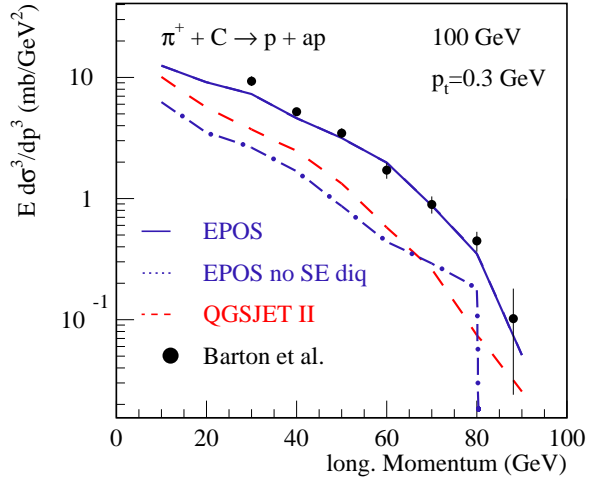}\end{center}
\caption{Model comparison: longitudinal momentum distributions of pion carbon 
collisions at 100 GeV from \epo~with (full) or without (dashed-dotted)
sting-end diquarks and QGSJETII (dashed) compared to data~\cite{barton}.}
\label{fig-se}
\end{figure}

Thus, besides the three valence quarks,
each remnant has additionally quarks and anti-quarks to compensate
the flavors of the string ends, as shown in fig.~\ref{nexus1}~c). 
According to its number of quarks and anti-quarks, to the phase space, and to
an excitation probability, a remnant decays into mesons and/or (anti)baryons \cite{nex-bar}. 
Furthermore, this process leads
to a baryon stopping phenomenon in which the baryon number 
can be transfered from the remnant to the string ends (for instance 
in~\ref{nexus1}~d), depending
on the process, the $3\overline{\mathrm{q}}+3\mathrm{q}$ can be
seen as 3 mesons or a baryon-antibaryon pair). 

In case of meson projectile, this
kind of diquark pair production at the string ends leads to an increase of the
(anti)baryon production in the forward production in agreement with low energy
pion-nucleus data~\cite{barton} as shown fig.~\ref{fig-se}. Comparing to 
{\sc qgsjetII} model~\cite{qgsjetII} which do not have diquark as string ends
or using only  q$\overline{\mathrm{q}}$ as string end in \epo, we can clearly
see that this process is needed to reproduce experimental data.
As a consequence it 
is part of the larger number of muons in air shower simulations with \epo.

Energy momentum sharing and remnant treatment are the key points of the model
concerning air shower simulations because they directly influence the 
multiplicity and the inelasticity of the model.

\section{NEW DEVELOPMENTS}

With the start of the LHC era, it is now possible to develop some particular
physics point of the \epo~model, which will be possible to test with newly
available data.

\subsection{Hydrodynamical Evolution}

In \epo~2, a new tool has been developed for treating 
the hydrodynamical evolution (see \cite{hydro} for details and tests 
with AuAu data).

As we saw, in case of elementary reactions like proton proton scattering 
(at moderately relativistic energies),
hadron production is realized via string breaking, such that string
fragments are identified with hadrons. When it comes to heavy ion
collisions or very high energy proton-proton scattering, the procedure
has to be modified, since the density of strings will be so high that
they cannot possibly decay independently. For technical reasons, we
split each string into a sequence of string segments, at a given proper-time
$\tau_{0}$, corresponding to widths $\delta\alpha$ and $\delta\beta$
in the string parameter space. One
distinguishes between string segments in dense areas (more than some
critical density $\rho_{0}$ of segments per unit volume), from those
in low density areas. The high density areas are referred to as core,
the low density areas as corona \cite{corona}. String segments
with large transverse momentum (close to a kink) are excluded from
the core. Based on the four-momenta of infinitesimal string segments,
\begin{equation}
\delta p=\left\{ \frac{\partial X(\alpha,\beta)}{\partial\beta}\delta\alpha+\frac{\partial X(\alpha,\beta)}{\partial\alpha}\delta\beta\right\} ,\end{equation}
 one computes the energy-momentum
tensor and conserved currents. The corresponding energy density $\varepsilon(\tau_{0},\vec{x})$
and the flow velocity $\vec{v}(\tau_{0},\vec{x})$ serve as initial
conditions for the subsequent hydrodynamic evolutions, which is characterized by :
\begin{itemize}
\item consideration of the possibility to have a (moderate) initial collective
transverse flow;
\item event-by-event procedure, taking into the account the highly irregular
space structure of single events, being experimentally visible via
so-called ridge structures in two-particle correlations; 
\item use of an efficient code for solving the hydrodynamic equations in
3+1 dimensions, including the conservation of baryon number, strangeness,
and electric charge; 
\item employment of a realistic equation-of-state, compatible with lattice
gauge results -- with a cross-over transition from the hadronic to
the plasma phase; 
\item use of a complete hadron resonance table, making our calculations
compatible with the results from statistical models;
\item hadronic cascade procedure after hadronization from the thermal system
at an early stage.
\end{itemize}

\subsection{Diffraction}

In order to produce hard diffractive events as measured at Tevatron and
most likely at LHC in the near future, it was necessary to improve the way
of producing diffractive events in \epo~2. To get a consistent description
of low mass and high mass diffraction, an effective diffractive Pomeron is
used in addition to the soft and semi-hard Pomeron. This effective object
represents in reality all type of diagrams (soft and hard component) which
are not connected to at least one of the remnant, leaving the latter intact.
The triple Pomeron would be part of this object, but all higher orders too.
Since this object is very difficult to calculate explicitly at all orders
including proper energy conservation, we use a simple parameterization having
the same form as the usual Pomeron (see \cite{splitting}) and fixing the
parameters using the proton-proton single diffractive cross section and the
energy spectrum of leading protons. 

\subsection{Structure Function}

In order to test hard scattering at LHC (or Tevatron) using \epo, it is
important to be able to reproduce the structure function
$F_2$ as measured thanks to the HERA electron-proton collider.

\begin{figure}[ht]
\begin{center}\includegraphics[width=0.36\textwidth]{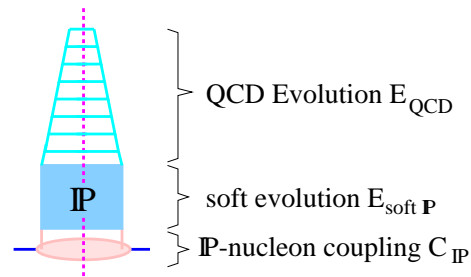}\end{center}
\caption{``Half-''Pomeron corresponding to the parton distribution function
 tested via the structure function $F_2$.}
\label{f2-diag}
\end{figure}

\begin{figure}[th]
\begin{center}\includegraphics[width=0.45\textwidth]{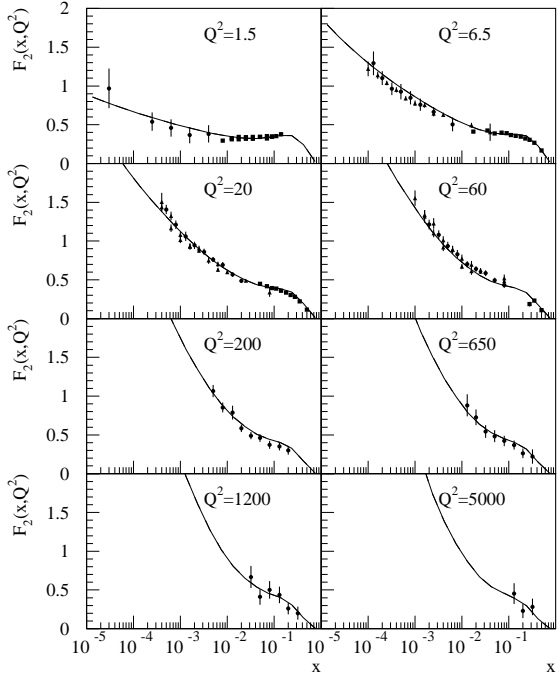}\end{center}
\caption{The structure function $F_2$ for different values of $Q^2$ together
with experimental data from H1~\cite{h1}, ZEUS~\cite{zeus} and NMC~\cite{nmc}.
\label{f2-data} }
\end{figure}

In \epo, $F_2$ corresponds to a ``half'' hard or semi-hard Pomeron as 
illustrated in fig.~\ref{f2-diag}. The cross section of a quark with a 
given $Q^2$ as a function of the momentum fraction $x$ can be calculated
using the convolution of a soft pre-evolution with the DGLAP equations for
the perturbative development of the parton ladder~\cite{nexus}. Taking into
 account the
effective corrections due to higher order diagrams in the connection between
the Pomeron and the nucleon (and fixed with
proton-proton cross section), we obtain the structure function as shown in 
fig.~\ref{f2-data}. 

\subsection{Baryon production}

In \epo~2, not only the soft Pomerons can have a diquark as string-end,
but the semi-hard Pomerons are now treated the same way. It will increase 
the production of (anti-)baryons
in the forward region compared to the previous version of \epo. This may
be seen in the number of muons produced in air shower simulations.

Furthermore, to take into account the property of different jet types, the 
string tension used for the string fragmentation now depends  on the initial
partons. Jets coming from quarks or from gluons will not produce the same
ratio of (anti-)baryons or strange particles over pions, and it will change
the energy evolution of such ratios.

\section{PRELIMINARY RESULTS AT LHC AND OUTLOOK}

\begin{figure}[ht]
\begin{center}\includegraphics[width=0.36\textwidth, angle=-90]{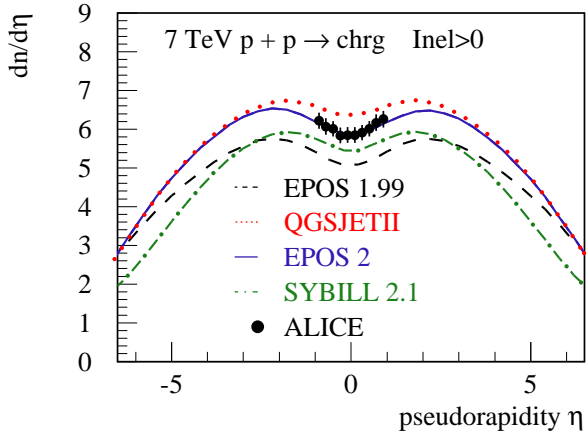}\end{center}
\caption{ Pseudorapidity distribution in $pp$ scattering at
7 TeV (INEL$>$0 trigger), compared to ALICE data (points) for different models :
results from \epo~2 (full) together with results from \epo~1.99 
(dashed), {\sc qgsjetII} (dotted) and Sibyll 2.1 (dashed-dotted).}
\label{fig-dndeta}
\end{figure}

The new treatment of the diffraction together with the better calculation of
the structure function and the more accurate treatment of the 
hydrodynamical phase, provide
much better results in comparison with LHC data than with \epo~1.99. We show
the pseudorapidity distribution at 7 TeV in fig.~\ref{fig-dndeta} as an 
example. \epo~2 results are compared to  \epo~1.99 and other models used for 
air shower simulations {\sc qgsjetII} and Sibyll~2.1~\cite{sibyll} and 
ALICE data~\cite{alice}.

More detailed results can be found in \cite{epos2pp,epos2ridge}, especially
concerning the effect of the hydrodynamical evolution on correlations between
secondary particles (Bose-Einstein correlation, ridge).

For the moment \epo~2 can only be used for minimum bias physics, but in the 
future it will be possible to select special class of hard events to study
specific channels and underlying events~\cite{porteboeuf}. 

\bigskip 

\end{document}